\documentclass[%
 reprint,
 amsmath,amssymb,
 aps,
floatfix,
]{revtex4-1}

\usepackage{graphicx}% Include figure files
\usepackage{dcolumn}% Align table columns on decimal point
\usepackage{bm}% bold math
\usepackage{siunitx}
\usepackage{xcolor}

\graphicspath{{figs/}}

\begin{document}

\title{The first open channel for a yield-stress fluid in complex porous media}

% Authors
\author{Dimitrios Fraggedakis}
\thanks{Corresponding author - email: dimfraged@gmail.com}
\affiliation{
   Department of Chemical Engineering,\\
   Massachusetts Institute of Technology,\\
   Cambridge, MA 02139 USA
}
\author{Emad Chaparian}
\thanks{email: emad@math.ubc.ca}
\affiliation{
   Mathematics Department, University of British Columbia,\\
   Vancouver, BC, V6T 1Z2, Canada
}
\author{Outi Tammisola}
\thanks{email: outi@mech.kth.se}
\affiliation{
   FLOW, Department of Engineering Mechanics,\\
   KTH Royal Institute of Technology,\\
   SE-10044 Stockholm, Sweden
}

\date{\today}

\begin{abstract}
The prediction of the first fluidized path of a yield-stress fluid in complex porous media is a challenging yet an important task to understand the fundamentals of fluid flow in several industrial and biological processes. In most cases, the conditions that open this first path are known either through experiments or expensive computations. Here, we present a simple network model to predict the first open channel for a yield-stress fluid in a complex porous medium. For porous media made of non-overlapping disks, we find that the pressure drop required to open the first channel for given yield stress depends on both the relative disks size to the macroscopic length of the system and the packing fraction. We also report the statistics on the arc-length of the first open path. Finally, we discuss the implication of our results on the design of porous media used in energy storage applications.
\end{abstract}

\maketitle

\section{\label{sec:introduction}Introduction}

Fluid flow in porous media is studied for more than a century due to its high relevance to several engineering applications such as enhanced oil recovery~\cite{green1998enhanced,farajzadeh2012foam,fraggedakis2015flow,sahimi2011flow}, filtration and separation~\cite{herzig1970flow,tien1979advances,jaisi2008transport}, fermentation~\cite{pandey2003solid,aufrecht2019pore}, soil sequestration~\cite{schlesinger1999carbon}, energy storage~\cite{duduta2011semi,sun2019hierarchical}, and food processing~\cite{greenkorn1983flow}. In most cases, the fluids involved in these applications exhibit yield stress/viscoplastic behavior~\cite{bonn2009yield,balmforth2014yielding}. Therefore, understanding the conditions -- critical pressure drop and/or stresses -- that lead to fluidization of yield-stress fluids in porous media can help boost the efficiency and lower the operational cost of several industrial applications.

In pressure-driven flows, the critical pressure drop $\Delta P_c$ required to fluidize the yield-stress fluid and open the first channel~\cite{chen2005flow,hewitt2016obstructed,waisbord2019anomalous} depends on the heterogeneous geometric characteristics and porosity $1-\phi$ of the porous medium, where $\phi$ is the volume fraction of the solid phase~\cite{talon2013geometry,bauer2019experimental,chaparian2020complex}. Therefore, it is crucial to understand the relation between the yielding conditions and the structure of the porous medium, which will lead to predictive models for both the first open channel and $\Delta P_c$.

The classical way to study yield-stress fluids is by solving the Cauchy momentum equations using viscoplastic constitutive relations, such as the Bingham and Herschel–Bulkley models~\cite{bird1987dynamics,huilgol2015fluid,saramito2016complex}. More recently, though, there is an increasing trend on using elastoviscoplastic~\cite{saramito2007new} and kinematic hardening~\cite{dimitriou2019canonical} models that originate from continuum mechanics~\cite{gurtin2010mechanics,anand2020continuum}. The yield stress behavior, however, leads to an ill-defined problem that does not describe the stress distribution within the unyielded regions of the fluid~\cite{balmforth2014yielding,saramito2017progress}. Common ways to resolve this problem is by using either optimization-~\cite{hestenes1969multiplier,powell1978algorithms,glowinski2008lectures} or regularization-based methods~\cite{papanastasiou1987flows,frigaard2005usage}. The former are accurate on predicting the yielded/unyielded boundaries and the flow field~\cite{dimakopoulos2013steady}, however, they are computationally expensive~\cite{saramito2016damped} for conditions nearby the yield limit. Although the latter reduce the computational cost, they introduce non-physical parameters~\cite{tsamopoulos2008steady,dimakopoulos2018pal} that lead to non-physical solutions, incorrect location of yield/unyield boundaries, and inaccurate yield limits~\cite{mitsoulis2017numerical,frigaard2005usage}. Here, we are interested in determining the statistics of the critical $\Delta P_c$ for a yield-stress fluid in complex porous media. Thus, we need to use models that can predict accurately and efficiently $\Delta P_c$ along with the first open channel.

Fluid flow in porous media is traditionally described through network models~\cite{fatt1956network} that represent the complex geometric characteristics of the domain with spherical pore throats and cylindrical edges~\cite{alim2017local,bryant1993network,blunt2001flow,blunt2013pore,stoop2019disorder}. In addition to their wide applicability in Newtonian fluids, network models have also been applied to describe $\Delta P_c$ and the flow behavior with respect to the applied pressure drop in yield-stress fluids~\cite{chen2005flow,frigaard2017bingham,liu2019darcy,talon2020effective}. When the relation between the local flow rate and the pressure drop is known, the network representation allows for the use of graph theoretic tools~\cite{kharabaf1997invasion,chen2005flow,balhoff2012numerical,liu2019darcy} to quickly evaluate $\Delta P_c$ and the flow response of the system. In general, though, the results of network viscoplastic models in complex porous media have been rarely compared and validated against those produced by solving the full fluid problem, and thus the conditions of their validity/applicability are unknown. 

The goal of the present work is to predict the first open channel for a yield-stress fluid in a complex porous medium along with the critical applied pressure drop required to open it. We develop a simple network model based on realistic porous media configurations, and use graph theoretical tools to study the statistics of yielding conditions in terms of the medium porosity. We validate our results against reported pressure-driven simulations of Bingham fluids in porous media. Finally, we discuss the relevance of our study to applications such as enhanced oil recovery and propose possible extensions.

\section{\label{sec:modeling}Theory}

\begin{figure*}[!ht]
    \centering
    \hspace{0.08in}\includegraphics[width=0.7\textwidth]{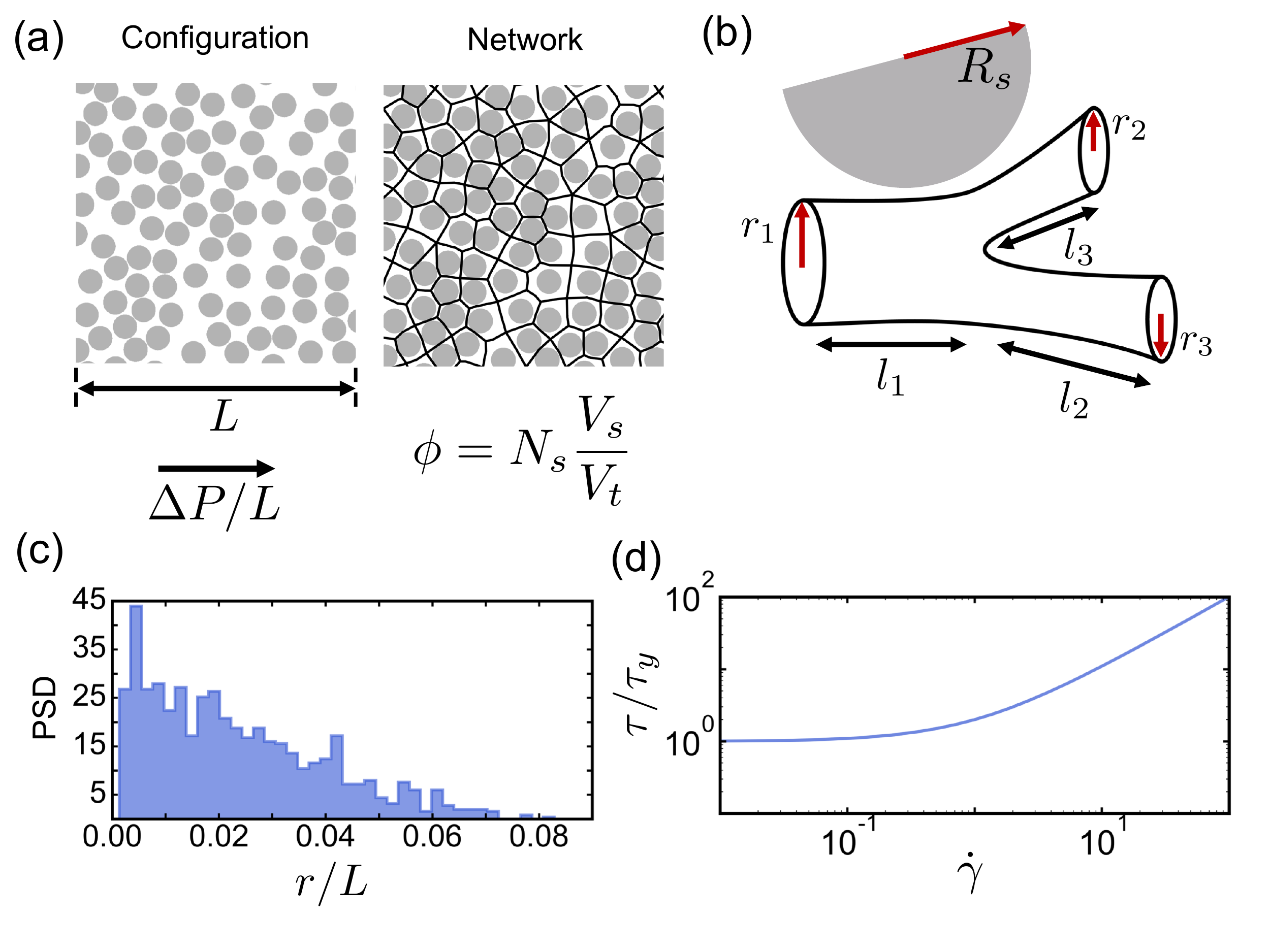}
    \caption{(a) Typical configuration of a porous medium of length $L$ and porosity $1-\phi$ that is made of non-overlapping monodisperse spheres of radius $R_s$, and its network representation. Across the domain, a macroscopic pressure gradient $\Delta P/L$ is applied to fluidize the yield-stress fluid. (b) Schematic of three network edges with different radii $r_i$ and lengths $l_i$ ($i=1,2,3$ ). The network representation includes the local geometric characteristics of the complex porous medium structure. (c) Characteristic pore-size distribution for the network shown in (a) as derived from the network model. (d) Typical shear stress response $\tau$ as a function of the applied shear rate $\dot{\gamma}$ of a viscoplastic fluid with yield stress $\tau_y$.}
    \label{fig:system}
\end{figure*}

\subsection{Network topology}
We are interested in the construction of realistic network models that capture the complex morphology of real porous media. The main scope of our work is to understand the statistics on the critical conditions that lead to fluidization in terms of the porosity $1-\phi$ and topological characteristics of the medium. 

To first approximation, we assume a porous medium that consists of monodisperse non-overlapping spheres/disks of radius $R_s$ as shown in Fig.~\ref{fig:system}(a). It is apparent that the structure of void space depends on the solid volume fraction defined as $\phi=N_sV_s/V_t$, where $N_s$ is the total number of spheres, $V_s$ is the volume of an individual sphere and $V_t$ the total volume of the system. We can use the given porous medium structure and create the network representation shown in the right panel of Fig.~\ref{fig:system}(a).

The network consists of nodes and edges that span the entire medium, where its complex topological characteristics are encoded on the connectivity between them~\cite{gostick2017versatile,khan2019dual}. Additionally, the local geometric characteristics of the porous medium are included on the length $l_i$ and radius $r_i$ of each individual edge, Fig.~\ref{fig:system}(b). For demonstration, we show in Fig.~\ref{fig:system}(c) the pore size distribution for the configuration of Fig.~\ref{fig:system}(a). Details on the generation of porous media with monodisperse non-overlaping sphere, the construction of the network representation, and the choice of $r_i$ and $l_i$ are given in the Appendix of the paper.

\subsection{Yield-stress fluid in a network}

Yield-stress fluids are characterized by their solid-liquid transition when the Euclidean norm of the stress field exceeds the value of yield stress (Von Mises criterion)~\cite{gurtin2010mechanics,hill1998mathematical}. The typical shear stress response in simple shear flow is shown in Fig.~\ref{fig:system}(d), where for $\dot{\gamma}\rightarrow0$ the shear stress reaches its critical value $\tau\rightarrow\tau_y$. The most common constitutive relations used to describe yield-stress fluids are the Bingham and the Herschel-Bulkley models~\cite{bird1987dynamics}. Both of them, though, have the same behavior for $\dot{\gamma}\rightarrow0$. Therefore, it is sufficient to discuss only the Bingham model for a porous medium to understand the connection between the yielding conditions to the geometric and topological characteristics of the network.

For pressure-driven flows, the local flow rate $q_i$ of edge $i$ is described in terms of the local geometric properties $r_i,l_i$, the local pressure drop along the edge $\Delta P_i$, and the yield stress $\tau_y$ of the fluid as~\cite{bird1987dynamics,liu2019darcy,frigaard2019background}
\begin{equation}
\label{eq:qflow}
q_{i} = \left\{ 
\begin{array}{cc}
\frac{r_i^4}{l_i} \left(\Delta P_{i}-\frac{\tau_y l_i}{r_i}\right) &\text{for} \,\, \Delta P_{i}>\frac{\tau_y l_i}{r_i}\\
 0   &\text{ for} \,\, \left\lvert\Delta P_{i}\right\rvert < \frac{\tau_y l_i}{r_i} \\
\frac{r_i^4}{l_i} \left(\Delta P_{i}+\frac{\tau_y l_i}{r_i}\right) &\text{ for} \,\, \Delta P_{i}<-\frac{\tau_y l_i}{r_i}\\
\end{array} 
\right.
\end{equation}
Near the no-flow limit $q_i\rightarrow0$, we see from Eq.~\ref{eq:qflow} that $\Delta P_i\rightarrow \tau_yl_i/r_i$, and thus smaller in radius or longer in length channels require larger applied pressure drop to yield. Across the first open channel, we can calculate the total pressure drop across the medium to be $\Delta P \equiv \sum_{i=1}^N \Delta P_i = \tau_y\sum_{i=1}^N l_i/r_i$, where $N$ is the total number of edges across the path. From this expression, we can see that the connectivity between the edges determines the first open channel in a real porous medium, and it corresponds to the path of `least resistance'. Thus, the problem of finding $\Delta P_c$ can be formulated as finding the path of the minimum pressure drop as follows~\cite{liu2019darcy}
\begin{equation}
\label{eq:least_res}
    \frac{\Delta P_c}{\tau_y} = 
    \min_{C \in \mathcal{C}_{\text{in-out}}}
    \sum_{i=1}^N \frac{l_i}{r_i}
\end{equation}
where $\mathcal{C}_{\text{in-out}}$ is the set of all paths between the corresponding boundaries. 

The problem of Eq.~\ref{eq:least_res} satisfies the principle of minimum dissipation rate and is valid near equilibrium~\cite{kondepudi2014modern}. In particular, the entropy production for a pressure-driven flow is $\sigma_D=q\Delta P$~\cite{de2013non}. Thus, for conditions near the solid-liquid transition where $q\rightarrow0^+$, the minimum pressure drop path also minimizes $\sigma_D$.

To solve Eq.~\ref{eq:least_res}, we transform the generated network into a graph with edges that have weights equal to $l_i/r_i$ and use the Dijkstra method~\cite{dijkstra1959note} for directed graphs to determine the first open channel. This method is known to scale quadratically with the path length~\cite{west1996introduction,bollobas2013modern}, and therefore for complex domains that lead to larger number of edges the computational cost increases. For a single porous medium configuration, however, the overall computational time to determine the first open channel is much lower (seconds to minutes) than that required to solve the full fluid flow problem using optimization methods (days to weeks)~\cite{dimakopoulos2018pal,chaparian2020complex}. 

\section{\label{sec:results}Results}

\begin{figure}[!ht]
    \centering
    \hspace{0.08in}\includegraphics[width=0.5\textwidth]{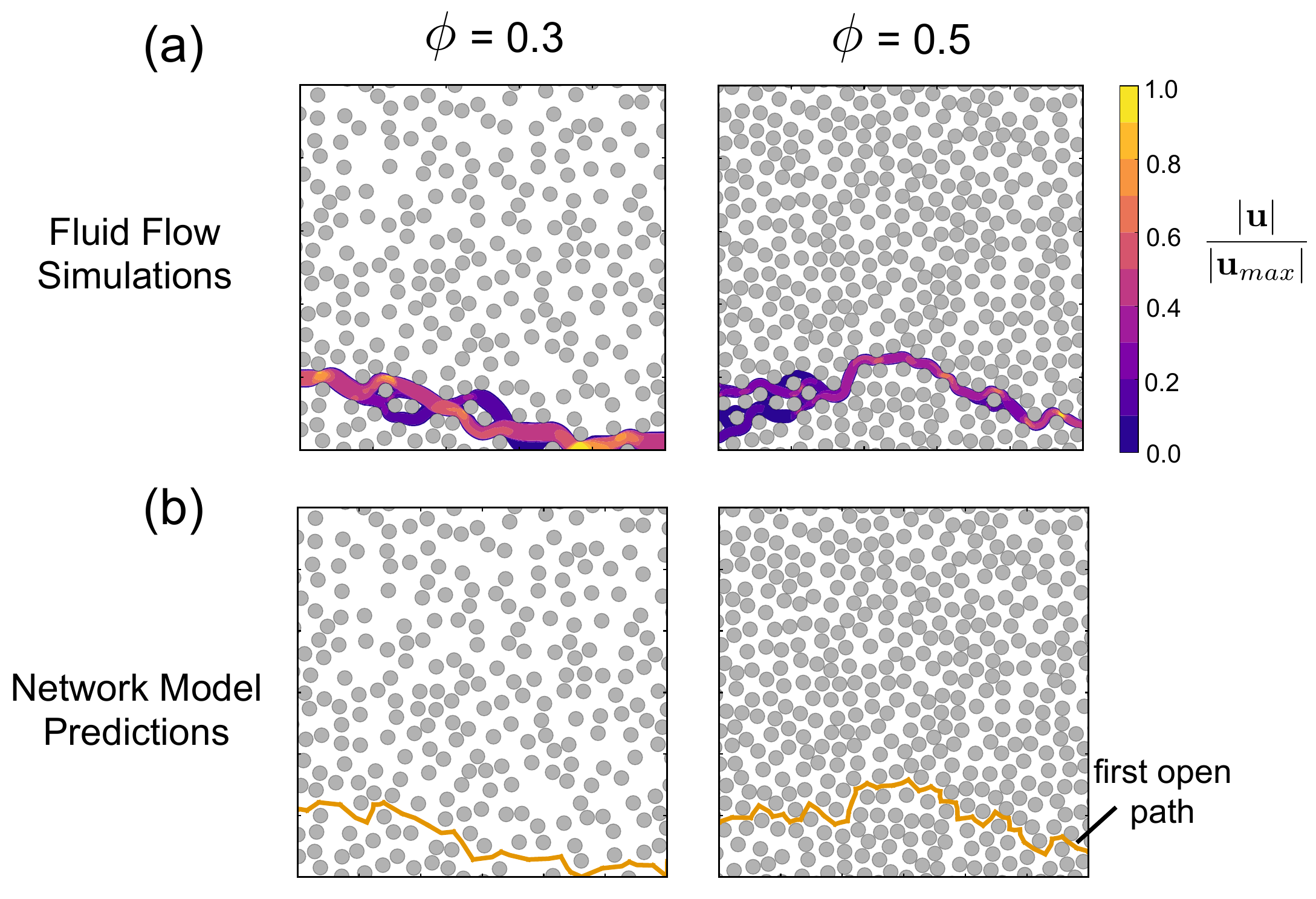}
    \caption{Validation of the network model for the first open channel against simulations for a pressure-driven Bingham fluid in a complex porous medium with $R_s/L=0.02$. (a) Simulation results of the open pathway for conditions near the critical pressure drop $\Delta P_c$. The contour plot shows the magnitude of the local velocity, normalized with the maximum velocity across the channel. (b) Network model predictions for the first open channel. The cases of $\phi=0.3$ and $\phi=0.5$ are examined. It is clear that both the full Bingham fluid simulation and the network model predict the same location for the first open channel.}
    \label{fig:validation}
\end{figure}

\subsection{The first open channel}
When the applied pressure drop approaches the critical value $\Delta P \rightarrow \Delta P_c^+$, there exists a single open channel across the entire medium. Here, we test the validity of the proposed approach to determine the first open channel nearby when the solid-liquid transition occurs. For comparison, we solve the full flow field under pressure-driven conditions for a yield-stress fluid for the complex porous media shown in Fig.~\ref{fig:validation}. We consider the cases of $\phi=0.3$ and $\phi=0.5$, respectively. All the lengths are normalized with the macroscopic length of the system $L$ and also $R_s/L=0.02$. For simplicity, we use two-dimensional porous media, however, our approach is general and does not depend on the dimensionality of the problem.

Figure~\ref{fig:validation}(a) shows the normalized velocity magnitude that results from the solution of the Cauchy momentum equation for a Bingham fluid~\cite{chaparian2020complex}. Also, Fig.~\ref{fig:validation}(b) depicts the predictions for the first open channel after solving the minimization problem of Eq.~\ref{eq:least_res}. In both cases, the network model is able to reproduce the results of the fluid problem for the first open path.

Notably, in both cases of Fig.~\ref{fig:validation}(a), where the full fluid flow problem is solved, we can see the existence of additional paths other than the one predicted by the network model. We justify this observation based on the fact that the fluid flow simulations are performed at extremely large, yet finite non-dimensional yield stress (i.e.~Bingham number; see \cite{chaparian2020complex}). Indeed, we speculate that for simulations with $q\rightarrow0^+$ (i.e.~infinite Bingham number), the secondary paths will eventually close and only the predicted one by the network model will be present. However, as it is clear from the Fig.~\ref{fig:validation}(a), flow rate in those secondary paths are negligible and does not contribute to the leading order of the resistance.

\subsection{Critical pressure drop $\Delta P_c$ and its statistics in complex porous media}

\begin{figure*}[!ht]
    \centering
    \hspace{0.08in}\includegraphics[width=0.7\textwidth]{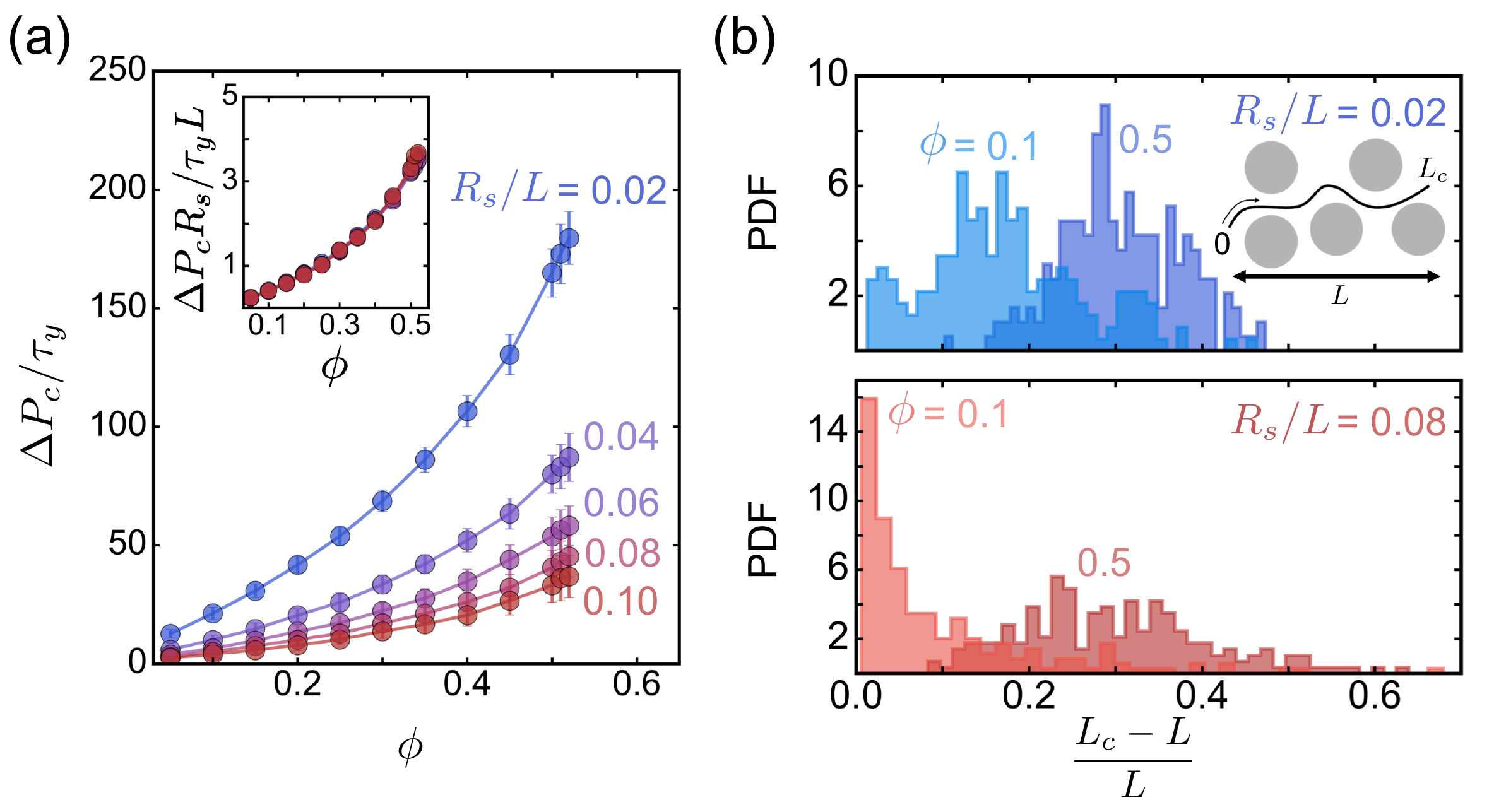}
    \caption{Statistics on the predictions of the network model for 500 realizations per value of $\phi$. (a) Critical pressure drop $\Delta P_c$ as a function of the volume fraction $\phi$ for different ratio of $R_s/L$. $\Delta P_c$ is normalized with the yield stress of the fluid $\tau_y$. The error bars indicate the variance around the mean value. Inset -- Scaled critical pressure drop $\Delta P_c R_s/\tau_y L$ in terms of $\phi$. For non-overlapping spheres, the results for different $R_s/L$ collapse in a master curve. (b) Probability density distributions for the normalized arc-length of the first open channel $L_c/L-1$. The cases of $R_s/L=0.02$ and $R_s/L=0.08$ are shown, respectively, for $\phi=0.1$ and $\phi=0.5$. In general, increasing $\phi$ leads to increase of $\Delta P_c$ required to open the first channel, but also leads to a wide distribution of arc-lengths, which provide large uncertainty on the critical macroscopic pressure gradient $\Delta P_c/L_c$ required to yield the fluid in the porous medium.}
    \label{fig:result}
\end{figure*}

In addition to the first open path, the network model can predict the normalized critical pressure $\Delta P_c$ required to yield the fluid in the porous medium. 

\begin{table}[h]
\begin{ruledtabular}
\begin{tabular}{ c c c c}
\textrm{$\phi$}&
\textrm{$\frac{\Delta P_c}{\tau_y}$ Network}&
\textrm{$\frac{\Delta P_c}{\tau_y}$ Simulations}&
\textrm{Rel. Error $\%$}\\
\colrule
$0.1$ & $22.16$ & $22.63$ & $1.87$\\
$0.3$ & $61$ & $65.2$ & $6.43$\\
$0.5$ & $142.16$ & $145.28$ & $2.05$\\
\end{tabular}
\end{ruledtabular}
\caption{\label{tab:table_DPc}%
Predicted normalized critical pressure drop $\Delta P_c$ required to open the first open channel for the configurations shown in Fig.~\ref{fig:validation}. We show both the predictions of the network model and the full simulation results, along with the relative error $\frac{\Delta P_{c}^{\text{sim}}-\Delta P_{c}^{\text{net}}}{\Delta P_{c}^{\text{sim}}}$. The network model is adequate on predicting both the first open channel and the critical pressure drop required to open it.
}
\end{table}

In table~\ref{tab:table_DPc} we show the predictions of the normalized critical pressure drop $\Delta P_c/\tau_y$ for both the network model and the full simulations. For $\phi=0.3$ and $0.5$ we consider the configurations shown in Fig.~\ref{fig:validation}. In all cases, it is clear that the relative difference of $\Delta P_c$ between the two models never exceeds $6.5\%$. The negligible difference can be justified by the fact that discussed above about the secondary paths, or/and the simplification of the geometric characteristics of the medium by its network representation. We conclude, though, that network-based models are adequate to predict both the first open path and the critical pressure drop required to open it. 

The validation of the model allows us to predict $\Delta P_c$ as a function of $\phi$ and the geometric characteristics of the system. For a porous medium made of non-overlapping disks, we control the microstructure characteristics by changing the ratio $R_s/L$. For each combination of $\phi$ and $R_s/L$, we generate 500 realizations for gathering the statistics of $\Delta P_c$. 

Figure~\ref{fig:result}(a) shows the normalized critical pressure drop $\Delta P_c/\tau_y$ in terms of $\phi$. The colored lines indicate different value of $R_s/L\in[0.02,0.1]$, while the error bars correspond to the variance of the statistical sample. For all $R_s/L$ the normalized pressure drop increases with increasing $\phi$. This behavior is expected as the radius $r_i$ of each edge decreases monotonically as $\phi$ increases~\cite{torquato2002random}. Additionally, we observe that decreasing the ratio $R_s/L$ leads to increase in $\Delta P_c$, in qualitative agreement with experiments~\cite{waisbord2019anomalous}. The reason for this behavior can be explained by examining the arc-length (defined as $L_c=\sum_i^N l_i$) for the first open channel. 

Figure~\ref{fig:result}(b) illustrates the histogram of $L_c$ for $\phi=0.3$ and $0.5$ as well as $R_s/L=0.02$ and $0.08$. In both cases we observe that increasing solid volume fraction leads to an increase in the total length of the first open path. While large $\phi$ results in almost similar behavior for both $R_s/L$, it is apparent that decreasing the size of the spheres results in more tortuous paths, even for low solid volume fraction. 

Dimensional analysis on Eq.~\ref{eq:least_res} indicates that $\Delta P_c$ follows a simple scaling with $R_s/L$. In particular, by rescaling the local edge radius $r_i$ with the sphere radius $R_s$, we find $\Delta \widetilde{P}_c=\Delta P_c R_s/\tau_y L = \sum_{i=1}^N \left(l_i/L\right)\left(R_s/r_i\right)$. The inset in Fig.~\ref{fig:result}(a) shows the rescaled form of the critical pressure drop for all $R_s/L$, were for non-overlapping disks, a master curve exists for all the examined values of $\phi$. The dependence of $\Delta \widetilde{P}_c$ with $\phi$, the effect of the pore microstructure, as well as the physical mechanism for the behavior of the tortuosity $\tau=L_c/L$ will be presented in a future work.

\section{\label{sec:discussion}Discussion}

We find that network models provide an efficient and accurate way to model the fluidization conditions in porous media. They are also accurate in locating the first open channel, which is equivalent to the path of least resistance through the entire medium.

Our results on the normalized critical pressure drop $\Delta P_c/\tau_y$ can be used to design porous media systems with the desired flow properties. In applications such as semi-solid flow batteries~\cite{duduta2011semi}, it is critical to keep the contact between the active material (electrode particles) and the conductive wiring (carbon nanoparticle network) intact during operation, otherwise there is a significant energy loss during cycling of the battery~\cite{solomon2018enhancing,wei2015biphasic}. This can be achieved by immersing the active material and the electronically conductive agent in yield-stress fluids like Carbopol~\cite{zhu2020high}. Therefore, we can use the predicted $\Delta P_c/\tau_y$ to determine the size of the active particles to optimize the design and operation of semi-solid electrodes.

The present model can also provide insights on the design of porous media. By taking advantage of the computational efficiency of the proposed network model, we can perform on-the-fly optimization to construct porous media with optimal mixing and transport properties~\cite{lester2013chaotic,kirkegaard2020optimal}. Such ideas have recently been implemented in elastic networks with optimal phonon band structures~\cite{ronellenfitsch2019inverse,ronellenfitsch2019chiral}, and we believe they can also be used for designing porous media immersed in yield-stress fluids.

Given their inherent node/edge structure, network models are fairly simple to be analyzed using graph theoretical tools. The unique property of yield-stress fluid, namely the fluidization conditions, allows us to use algorithms that can find the minimum resistance pathways with minimal efforts. In coarse-grained domains, however, where the microscopic geometric irregularities are encoded in the heterogeneous `permeability' tensor~\cite{hewitt2016obstructed}, graph theory tools might not be the most suitable ones. An alternative way to calculate the first open channel in a continuum with spatially variable properties is through methodologies used in physical chemistry to identify reaction pathways~\cite{henkelman2000climbing,swenson2018openpathsampling}. There, the first open channel corresponds to the path that passes through the minimum energy barrier, namely the transition state point.

The present model considers only the case of viscoplastic materials and disregards the fluid elasticity~\cite{saramito2007new,cheddadi2012steady,fraggedakis2016yielding1,fraggedakis2016yielding2} prior to yielding. Therefore, further analysis is required to connect the yield criterion to the elastic modulus of the fluid, which can provide insights for engineering both the porous medium and the fluid itself. Additionally, the yielding and/or stoppage conditions might be further affected by possible thixotropic~\cite{mewis2012colloidal} and kinematic hardening~\cite{dimitriou2019canonical,gurtin2010mechanics} phenomena.

\section{\label{sec:conc}Summary}
In this work, we presented a network model for yield-stress fluids in porous media that describe the effects of porosity $1-\phi$ and microstructure properties in complex geometries. We demonstrated the capabilities of the model to predict the critical applied pressure drop required to open the first channel in the medium. Also, we compared our results to direct numerical simulations of the full fluid problem for Bingham fluids and we showed the accuracy and computational efficiency of the network-based models to solid-liquid transition. Finally, we discussed the implications of our model on the optimization and design of porous media.

\section*{Contributions}
D.F. conceptualized, designed, and performed the analysis in the present study. E.C. and O.T. provided the fluid flow simulation results. D.F. wrote the manuscript. All authors contributed to the final manuscript. 

\section*{Acknowledgment}
D.F. (aka dfrag) wants to thank T. Zhou and M. Mirzadeh for insightful discussions related to the validity of the network model. The authors declare no competing interests.

\section*{\label{sec:appendix}Appendix}
\subsection*{Porous medium and network generation}
For the generation of porous media that consist of non-overlapping disks, we implemented the random sequential addition (RSA) algorithm~\cite{zhang2013precise,cule1999generating,torquato2006random}. The procedure described in~\cite{torquato2006random} allows for the fast generation of randomly packed disks with the desired volume fractions. Due to the constraint of non-overlapping disks, all generated microstructures never exceed $\phi=0.52$ in two dimensions. 

For the generation of the network model, we implemented the maximal ball algorithm as described in~\cite{silin2006pore,dong2009pore,al2007network}. The procedure allows us to get the radius $r_i$ and length $l_i$ for each edge. The maximal ball algorithm represents `fits' a circle/sphere within each pore~\cite{alim2017local}, its radius of which represents the $r_i$ we use in Eq.~\ref{eq:least_res}. Other choices of $r_i$ can be considered (e.g. equivalent radius etc.), however, our choice for the local radius works well in predicting both the first open channel and the critical pressure drop required to open it. The generated network was represented by a graph with vertices $V$ and edges $E$ using the open source library NetworkX~\cite{hagberg2008exploring}.

Details on the numerical simulation of the fluid flow problem shown in Fig.~\ref{fig:validation}(a) can be found in~\cite{chaparian2020complex}.

\bibliography{literature}% Produces the bibliography via BibTeX.

\end{document}